\documentclass[prb,twocolumn,showpacs]{revtex4}
\usepackage{graphicx}

\begin{document}

\title{Hole-pair symmetry and excitations in the strong-coupling extended
$t-J_z$ model: competition between $d$-wave and $p$-wave}


\author{R. M. Fye}
\affiliation{Department of Physics and Astronomy, University of New 
Mexico, Albuquerque, NM 87131}
\author{G. B. Martins}
\email[Author to whom correspondence should be addressed at: 
]{martins@magnet.fsu.edu}
\author{E. Dagotto}
\affiliation{Department of Physics and National High Magnetic Field Lab, 
Florida State University, Tallahassee, FL
32306}

  \begin{abstract}
  We analytically calculate the ground state
  pairing symmetry and excitation
  spectra of two holes doped into the half-filled
  $t- t^\prime -t^{\prime \prime}-J_z $ model in the strong-coupling
  limit ($J_z >> |t|, |t^\prime|, |t^{\prime \prime}|$). In leading order, this reduces to 
  the $t^\prime -t^{\prime \prime}-J_z $ model, where there are
  regions of $d$-wave,
  $s$-wave, and (degenerate) $p$-wave symmetry.
  We find that the
  $t-J_z$ model maps in lowest order onto the
  $t^\prime -t^{\prime \prime}-J_z $ model on the boundary between
  $d$ and $p$ symmetry, with a flat lower band in the pair excitation 
spectrum.
  In higher order, $d$-wave symmetry
  is selected from the lower pair band. However, we observe that
  the addition of the appropriate
  $t'<0$ and/or $t''>0$, the signs of $t^\prime$ and $t^{\prime\prime}$
  found in the hole-doped cuprates, could drive
  the hole-pair symmetry to $p$-wave, implying
  the possibility of competition between $p$-wave and $d$-wave pair ground 
states. (An added $t^\prime > 0$ and/or $t^{\prime \prime}  < 0$ generally tend 
to promote $d$-wave symmetry.)
  We perturbatively construct an extended quasi-pair for the $t-J_z$ model.
  In leading order, there are contributions from sites at a distance 
of $\sqrt{2}$
  lattice spacings apart; however, contributions
  from sites 2 lattice spacings apart, also
  of the same order, vanish identically.
  Finally, we compare our approach with analytic calculations for a 
$2\times2$ plaquette
  and  with existing numerical work, and discuss possible relevance to the physical 
parameter
  regime.

  \end{abstract}

  \maketitle

  \section{Introduction}

  In recent years a number of
  experiments, particularly phase-sensitive ones, have indicated that the 
pair symmetry
  of hole-doped
  cuprate superconductors is at least predominantly $d_{x^2 - 
y^2}$\cite{dexp1,annet,tsuei-kirtley}.
  Theoretical and
  numerical studies of the two-dimensional Hubbard, $t-J$, and related
  models have also suggested $d_{x^2 - y^2}$
  pairing\cite{dagottorev,schrieffer,reviews,scalrev,joynt,hlubina},
  and studies of Hubbard and $t-J$ models
  on $2\times2$ plaquettes have provided an
  intuitive picture of how $d$-wave symmetry might 
arise\cite{scaltrug,whitescal}.
  However, there are few rigorous theoretical results in this general area.

  Different experimental techniques have indicated that a pseudogap
  with the same symmetry as the superconducting gap
  persists above $T_c$
  in underdoped cuprates 
\cite{dexp1,pseudo1,pseudo2,preform,timusk-statt,tallon}.
  This, along with the short high-$T_c$ coherence
  length \cite{dagottorev},
  is qualitatively consistent with a strong-coupling picture,
  where pairs can preform at $T > T_c$ \cite{nan}.
  Numerical work has in addition
  suggested that the $t-J$ and $t-J_z$ models
  have many similar properties\cite{dagottorev,dagottop,tjz,hamer98}(see however Ref. \onlinecite{boninsegni}), and 
that
  the $t-J_z$ model may hence provide a suitable starting point for
  understanding $t-J$ behavior\cite{cumulant}.

  Reductions from CuO$_2$ three-band and similar models \cite{param}, as 
well as
  comparison with ARPES results for a single doped hole\cite{onefit},
  suggest that, besides a nearest-neighbor (NN) $t$, the next-NN (NNN) 
$t^{\prime}$
  and next-NNN (NNNN) $t^{\prime \prime}$ hoppings may also
  be substantial. Numerical calculations and theory
  have explored some of the qualitative effect that the sign
  of $t^{\prime}$ has upon hole-pairing\cite{onefitsps,whitescal1,martins1}.
  Given these above
  results, it is interesting to try to gain
  a better understanding of the pairing properties of the {\it extended} $t-J$ 
model.
  In addition, performing
  such a study analytically could help provide guidance to future
  numerical work aimed at exploring specific regions of the model's
  phase diagram.

  To that end, we consider in this paper two holes doped into the 
half-filled
  $t- t^\prime -t^{\prime \prime}-J_z$ model in the strong coupling limit
  ($J_z >> |t|, |t^\prime|, |t^{\prime \prime}|$).
  We calculate the symmetry of the hole pair in the ground state
  as well as the pair excitation spectrum. We do not explicitly
  consider the issues of
  phase separation or whether superconductivity actually occurs.
  We consider first the $t^\prime-J_z$ model,
  and show how singlet pairs
  can be constructed from our solutions.
  We next discuss the
  $t^\prime -t^{\prime \prime}-J_z $ model, and then the
  $t-J_z$ and
  $t-t^{\prime}-t^{\prime \prime} -J_z $ models.
  For the $t-J_z$ model, we perturbatively construct
  an extended quasi-pair.
  As a step towards exploring the
  range of validity of our approach,
  we compare with results
  for a 2$\times$2 plaquette and with numerical studies.
  Lastly, we discuss implications of our results
  for the physically relevant parameter regime,
  among which a phase competition scenario is one possibility.

  Specifically, we consider the Hamiltonian
  \begin{equation}
  H = H_0 + H_1 + H_2 + H_3,
  \label{h}
  \end{equation}
  where
  \begin{eqnarray}
  H_0 =
  J_z &\sum_{ x,y }& \{  ( S_{x,y}^{z} S_{x+1,y}^{z} +
  S_{x,y}^{z} S_{x,y+1}^{z} ) \nonumber\\
  && - {1 \over 4} ( n_{x,y} n_{x+1,y} + n_{x,y} n_{x,y+1} ) \},
  \label{h0}
  \end{eqnarray}
  \begin{eqnarray}
  H_1 =
  (-t) &\sum_{x,y,\sigma}& \{
  ( {\tilde c}_{x,y,\sigma}^\dagger
  {\tilde c}_{x+1,y,\sigma} + H.c.) \nonumber\\
  && + \,
  ( {\tilde c}_{x,y,\sigma}^\dagger
  {\tilde c}_{x,y+1,\sigma} + H.c.) \},
  \label{h1}
  \end{eqnarray}
  \begin{eqnarray}
  H_2 =
  (-t^\prime) &\sum_{x,y,\sigma}& \{
  ( {\tilde c}_{x,y,\sigma}^\dagger
  {\tilde c}_{x+1,y+1,\sigma} + H.c.) \nonumber\\
  && + \,
  ( {\tilde c}_{x,y,\sigma}^\dagger
  {\tilde c}_{x+1,y-1,\sigma} + H.c.) \},
  \label{h2}
  \end{eqnarray}
  and
  \begin{eqnarray}
  H_3 =
  (-t^{\prime\prime}) &\sum_{x,y,\sigma}& \{
  ( {\tilde c}_{x,y,\sigma}^\dagger
  {\tilde c}_{x+2,y,\sigma} + H.c.) \nonumber\\
  &&  + \,
  ( {\tilde c}_{x,y,\sigma}^\dagger
  {\tilde c}_{x,y+2,\sigma} + H.c.) \}.
  \label{h3}
  \end{eqnarray}
  Here, $x$ and $y$ denote the coordinates of an $L$x$L$ lattice
  with periodic boundary conditions
  and even $L$ (with, except for the $t^\prime-J_z$ model, $L>2$), 
and $\sigma = \pm 1 \,\, (\uparrow, \downarrow)$ refers to
  electron spin.
  ${\tilde c}_{x,y,\sigma} = c_{x,y,\sigma} ( 1 - n_{x,y,-\sigma} )$,
  enforcing the condition of no double occupancy.
  $S_{x,y}^{z} = {1 / 2} \,\, ( n_{x,y,\uparrow} - n_{x,y,\downarrow} )$
  and
  $n_{x,y} = n_{x,y,\uparrow} + n_{x,y,\downarrow}$.
  We do not explicitly
  consider here the spin-flip part of the magnetic interaction
  \begin{eqnarray}
  H_{\perp} =
  \Bigl( { {J_{\perp}} \over 2 } \Bigr)
  \sum_{ x,y }  \left\{ ( S_{x,y}^{+} S_{x+1,y}^{-} +
  S_{x,y}^{+} S_{x,y+1}^{-} ) + H.c. \right\},
  \label{jperp}
  \end{eqnarray}
  where $ S_{x,y}^{+} = c^{\dagger}_{x,y,\uparrow} c_{x,y,\downarrow}$
  and $ S_{x,y}^{-} = c^{\dagger}_{x,y,\downarrow} c_{x,y,\uparrow}$.
  (The full $t - t^{\prime} - t^{\prime \prime} - J$ model is recovered when
  $J_{\perp} = J_z$.)

  At half filling each site is occupied by exactly  one electron,
  and the doubly degenerate ground state of $H_0$
  is then that of a N\'eel
  antiferromagnet.
  We choose $|\Phi_{a}>$ to denote the state with electron spins
  $\sigma(x,y) = (-1)^{x+y}$ and $|\Phi_{b}>$ to denote the state
  with $\sigma(x,y) = (-1)^{x+y+1}$.
  We define the operator $a_{x,y} = c_{x,y,\sigma(x,y)}$ with
  $\sigma(x,y) = (-1)^{x+y}$,
  and the operator $b_{x,y} = c_{x,y,\sigma(x,y)}$ with
  $\sigma(x,y) = (-1)^{x+y+1}$
  Although our calculations and results are independent of the ordering
  convention chosen, we will denote for specificity
  \begin{equation}
  |\Phi_a> = (a^{\dagger}_{L,L}...a^{\dagger}_{1,L})...
    (a^{\dagger}_{L,2}...a^{\dagger}_{1,2})
    (a^{\dagger}_{L,1}...a^{\dagger}_{1,1})|0>,
  \label{gsa}
  \end{equation}
  where $|0\rangle$ is the state with no electrons,
  with an analogous definition for $|\Phi_b>$.

  We now dope the half-filled
  state $|\Phi_a>$
  with two holes and consider the strong-coupling
  limit ($J_z >> |t|, |t^\prime|, |t^{\prime \prime}|$). In this limit,
  there will be an energy cost of order $J_z$ if the two holes are
  {\it not} NN. Hence, to zeroth order,
  the (highly degenerate) two-hole ground state is spanned by the set of all
  NN hole pairs.
  We denote the state with a horizontal NN hole pair at sites
  $(x,y)$ and $(x+1,y)$ as
  \begin{equation}
  |h_{x,y}> = a_{x+1,y} a_{x,y} |\Phi_a>,
  \label{hxy}
  \end{equation}
  and the state with a vertical NN hole pair at sites $(x,y)$
  and $(x,y+1)$
  \begin{equation}
  |v_{x,y}> = a_{x,y+1} a_{x,y} |\Phi_a>.
  \label{vxy}
  \end{equation}
  The $|h_{x,y}>$'s and $|v_{x,y}>$'s provide a complete, orthonormal basis
  for the two-hole ground state of $H_0$ corresponding to $|\Phi_a>$.

  It costs an energy of order $J_z$ if one of the NN holes
  hops to a NN site through the hybridization matrix element $t$.
  However, there is no energy cost
  for hops corresponding to $t^{\prime}$ or $t^{\prime \prime}$, as
  long as the two holes remain NN after the hop.
  Thus, to lowest order in $1 / J_z$, it is only necessary
  to diagonalize the Hamiltonian $H_2 + H_3$ in the
  subspace spanned by the $|h_{x,y}>$'s and $|v_{x,y}>$'s; i.e., it is only
  necessary to consider the
  $t^\prime -t^{\prime \prime}-J_z $ model.
  We note that in this
  limit the $t^\prime -t^{\prime \prime}-J_z $ model becomes
  isomorphic to the strong-coupling limit of the
  antiferromagnetic van Hove model of ref. \onlinecite{afvh}.

  \section{$\lowercase{t}^{\prime}-J_{\lowercase{z}}$ Model}

  We consider first the $t^\prime -J_z $ model, involving only the
  $H_2$ (diagonal) hopping term.
  Defining
  \begin{equation}
  |h_{k_{x},k_{y}}> = {1 \over L}
  \sum_{ x,y } e^{-{ { 2 \pi i k_{x} x } \over L }}
  e^{-{ { 2 \pi i k_{y} y } \over L }} \, |h_{x,y}>
  \label{hk}
  \end{equation}
  and
  \begin{equation}
  |v_{k_{x},k_{y}}> = {1 \over L}
  \sum_{ x,y } e^{-{ { 2 \pi i k_{x} x } \over L }}
  e^{-{ { 2 \pi i k_{y} y } \over L }} \, |v_{x,y}>,
  \label{vk}
  \end{equation}
  with $k_{x},k_{y} = 0,1... \, L-1$,
  we obtain the lowest order wave functions
  \begin{eqnarray}
  |\psi^{\pm}_{k_{x},k_{y}}> &=&
  {1 \over {\sqrt 2}} \,
  \Bigl\{ e^{-{ { \pi i k_{x} x } \over L }}
  \, |h_{k_{x},k_{y}}> \nonumber\\
  && \, \pm \,\,  {\rm sgn}(t^{\prime}) \,
  e^{-{ { \pi i k_{y} y } \over L }}  \, |v_{k_{x},k_{y}}> \, \Bigr\}
  \label{tpwf}
  \end{eqnarray}
  with energies
\begin{equation}
  \epsilon^{\pm}_{k_{x},k_{y}} = \pm \, 4 \, |t^{\prime}|
    \> \sin \Bigl( { { \pi k_{x} } \over L } \Bigr)
    \> \sin \Bigl( { { \pi k_{y} } \over L } \Bigr).
\end{equation}

  Since
  $0 \le \sin (\pi k_{x} / L), \sin (\pi k_{y} / L) \le 1$,
  the minus sign gives the branch of lower energy.
  The lowest energy state $| \psi ^{(a)} _{0} >$,
  with energy $-4 \, |t^{\prime}|$, occurs when
  $k_{x} = L / 2$ and $k_{y} = L / 2$
  (i.e., $( \pi, \pi )$).
  Rewriting in terms of the $a_{x,y}$'s and neglecting overall
  phase factors, one obtains
  \begin{eqnarray}
  | \psi ^{(a)} _{0} > = {1 \over {L \sqrt 2}}
  &\sum_{ x,y }& (-1)^{x+y} \{ a_{x+1,y} a_{x,y} \nonumber\\
  && \, - \,\, {\rm sgn}(t^{\prime})\,  a_{x,y+1} a_{x,y} \} | \Phi_{a}>.
  \label{tpgs}
  \end{eqnarray}
  When $t^{\prime} > 0$ (sgn($t^{\prime}$) = 1),
  the sum over hole pair operators in Eq. \ref{tpgs}
  changes sign upon a 90 degree
  rotation around a lattice point, giving the pair $d$-wave symmetry
  (specifically, $d_{x^2 - y^2}$\cite{annet,scalrev,tsuei-kirtley}).
  When $t^{\prime} < 0$, there are no such sign changes,
  giving $s$-wave symmetry
  (specifically, extended-$s$\cite{annet,scalrev}).

  If one adds to Eq. \ref{tpgs}
  the appropriately-phased pair operator
  for two holes doped into the
  ground state $| \Phi_{b} >$, given in 
  \begin{eqnarray}
  | \psi ^{(b)} _{0} > = {1 \over {L \sqrt 2}}
  &\sum_{x,y}& (-1)^{x+y+1} \{ b_{x+1,y} b_{x,y} \nonumber\\
  && \, - {\rm sgn}(t^{\prime})\,  b_{x,y+1} b_{x,y} \} | \Phi_{b}> ,
  \label{tpbgs}
  \end{eqnarray}
  one obtains for $t^{\prime} > 0$ the usual 
  NN singlet $d_{x^2 - y^2}$
  pair operator
  \begin{eqnarray}
  \Delta_d= \nonumber \\
{1 \over {2L}}
  &\sum_{x,y}&
  \Bigl\{ \bigl( c_{x,y,\uparrow} c_{x+1,y,\downarrow} -
                 c_{x,y,\downarrow} c_{x+1,y,\uparrow} \bigr) \nonumber\\
     && - \bigl( c_{x,y,\uparrow} c_{x,y+1,\downarrow} -
                 c_{x,y,\downarrow} c_{x,y+1,\uparrow} \bigr) \Bigr\} ,
  \label{ccd}
  \end{eqnarray}
  with $t^{\prime} < 0$ giving the analogous singlet extended-$s$ operator
  \begin{eqnarray}
  \Delta_s= \nonumber \\
{1 \over {2L}}
  &\sum_{x,y}&
  \Bigl\{ \bigl( c_{x,y,\uparrow} c_{x+1,y,\downarrow} -
                 c_{x,y,\downarrow} c_{x+1,y,\uparrow} \bigr) \nonumber\\
    && + \bigl( c_{x,y,\uparrow} c_{x,y+1,\downarrow} -
                 c_{x,y,\downarrow} c_{x,y+1,\uparrow} \bigr) \Bigr\}.
  \label{ccs}
  \end{eqnarray}
  With different relative phases, one can also obtain types of $d$-wave
  or $s$-wave $m=0$ triplet pairs;
  because quantum spin fluctuations are not included in the $t-J_z$
  model, the cases cannot be differentiated at this
  level.

  One can better understand the dependence of the $t^{\prime}-J_z$ (and $t^{\prime}-J$) pair
  symmetry on sgn($t^{\prime}$) by considering phase transformations of
  electron creation and destruction operators. Specifically, we consider
  transformations of the form $c_j=e^{i\theta(j)}d_j$, where $j$ is some
  generalized coordinate referring to both orbital and spin and $\theta(j)$ 
is
  some function of $j$.

  First, as background, let $P$ denote some arbitrary product of
  electron creation and destruction operators, some of which operators
  may be the same, with coordinates referring to orthogonal states.
  For example, for a one-dimensional chain with one orbital per site, one
  could have 
$P=c^{\dagger}_{1\uparrow}c^{\dagger}_{4\downarrow}c_{1\uparrow}c^{\dagger}_{2\downarrow}
  c^{\dagger}_{1\uparrow}$.

  As previously, let $|0\rangle$ denote a state with no electrons. Then,
  $\langle0|P^{\dagger}P|0\rangle$ if it is nonzero reduces to
  \begin{eqnarray}
  \langle0|P^{\dagger}P|0\rangle =
  \langle0|\prod_{\{j\}}c_jc^{\dagger}_j|0\rangle
  = \langle0|\prod_{\{j\}}(1-n_j)|0\rangle
  \label{i1}
  \end{eqnarray}
  for some subset $\{j\}$ of the generalized $j$ coordinates. Hence, any
  phase transformation of the form $c_j=e^{i\theta(j)}d_j$ does not affect 
the
  value of $\langle0|P^{\dagger}P|0\rangle$.

  Now, consider $\langle0|P_l^{\dagger}P_{l^{\prime}}|0\rangle$, where 
$P_l$ and
  $P_{l^{\prime}}$ are two different products of creation and destruction 
operators. Each
  nonzero $\langle0|P_l^{\dagger}P_{l^{\prime}}|0\rangle$
  also reduces to the general form
  $\langle0|\prod_{\{j\}}c_jc^{\dagger}_j|0\rangle$, which is again 
unaffected by
  the transformations $c_j=e^{i\theta(j)}d_j$. Lastly, consider the 
expectation value of an
  operator $O=\sum_l a_l P_l$ in the (possibly unnormalized) state $\sum_l 
b_l P_l|0\rangle$,
  where the $a_l$ and $b_l$ are coefficients. Then,
  \begin{eqnarray}
  \langle O \rangle & = &
  {{\langle0|\sum_l 
b_l^*P^{\dagger}_lO\sum_{l^{\prime}}b_{l^{\prime}}P_{l^{\prime}}|0\rangle}\over
  {\langle0|\sum_l b_l^*P^{\dagger}_l\sum_{l^{\prime}} 
b_{l^{\prime}}P_{l^{\prime}}|0\rangle}} \nonumber\\
  & = & 
{{\sum_{l,l^{\prime},l^{\prime\prime}}a_{l^{\prime\prime}}b_l^*b_{l^{\prime}}
  \langle0|P^{\dagger}_lP_{l^{\prime\prime}}P_{l^{\prime}}|0\rangle} \over
  {\sum_{l,l^{\prime}} b_l^*b_{l^{\prime}} 
\langle0|P_{l}^{\dagger}P_{l{^\prime}}|0\rangle}}.
  \label{i3}
  \end{eqnarray}
  Again, each term of the numerator and denominator reduces to the form 
$\langle0|
  \prod_{\{j\}}c_jc^{\dagger}_j|0\rangle$, which is invariant under 
$c_j=e^{i\theta(j)}d_j$.
  Hence, an arbitrary phase transformation on electron creation and 
destruction operators
  has no effect on fermion operator expectation values in fermion states. 
Also, since an
  electron creation/destruction operator referring to a particular basis can 
always be written
  as a sum of operators referring to a different orthogonal basis, it is not 
even necessary
  that the creation/destruction operators under consideration refer to a 
particular
  orthogonal basis, as long as the phase transformations are consistent.

  Now, we wish to find a particular phase transformation which is equivalent 
to reversing
  the sign of $t^{\prime}$. Temporarily dropping the spin coordinate, we 
choose for simplicity
  a transformation of the form
  \begin{eqnarray}
  c_{x,y}=e^{i(\phi_xx+\phi_yy)}d_{x,y},
  \label{i4}
  \end{eqnarray}
  where $\phi_x$ and $\phi_y$ are constants. (We do not include a constant 
phase factor as
  it has no relevant effect.)
  We require:
  \begin{eqnarray}
  c^{\dagger}_{x+1,y+1}c_{x,y}+H.c.=\, 
(-1)[d^{\dagger}_{x+1,y+1}d_{x,y}+H.c.]
  \label{i5}
  \end{eqnarray}
  and
  \begin{eqnarray}
  c^{\dagger}_{x-1,y+1}c_{x,y}+H.c.=\, 
(-1)[d^{\dagger}_{x-1,y+1}d_{x,y}+H.c.].
  \label{i6}
  \end{eqnarray}
  Solving the previous equations, one can obtain
  \begin{eqnarray}
  \phi_x+\phi_y=(2p+1)\pi
  \label{i7}
  \end{eqnarray}
  and
  \begin{eqnarray}
  \phi_x-\phi_y=(2q+1)\pi
  \label{i8}
  \end{eqnarray}
  for arbitrary integers $p$ and $q$, of which solutions are 
  \begin{eqnarray}
  c_{x,y}=(-1)^{x}d_{x,y},
  \label{i9}
  \end{eqnarray}
  or
  \begin{eqnarray}
  c_{x,y}=(-1)^{y}d_{x,y}.
  \label{i10}
  \end{eqnarray}
  Restoring the spin coordinates leads to the four following
  transformations equivalent to changing the sign of $t^{\prime}$:
  \begin{eqnarray}
  c_{x,y,\uparrow}=(-1)^{x}d_{x,y,\uparrow}, \hspace{0.3cm} 
c_{x,y,\downarrow}=(-1)^{x}d_{x,y,\downarrow};
  \label{i11}
  \end{eqnarray}
  \begin{eqnarray}
  c_{x,y,\uparrow}=(-1)^{y}d_{x,y,\uparrow}, \hspace{0.3cm} 
c_{x,y,\downarrow}=(-1)^{y}d_{x,y,\downarrow};
  \label{i12}
  \end{eqnarray}
  \begin{eqnarray}
  c_{x,y,\uparrow}=(-1)^{x}d_{x,y,\uparrow}, \hspace{0.3cm} 
c_{x,y,\downarrow}=(-1)^{y}d_{x,y,\downarrow};
  \label{i13}
  \end{eqnarray}
  \begin{eqnarray}
  c_{x,y,\uparrow}=(-1)^{y}d_{x,y,\uparrow}, \hspace{0.3cm} 
c_{x,y,\downarrow}=(-1)^{x}d_{x,y,\downarrow}.
  \label{i14}
  \end{eqnarray}

  None of the above four transformations changes the sign of the $J_z$ term 
(Eq. \ref{h0}).
  However, the transformations of Eqs. \ref{i13} and \ref{i14} (as well as 
more general transformations not discussed here which are different on different 
sublattices) do not leave the $H_{\perp}$ of Eq. \ref{jperp}, and hence the 
spin-spin interaction of the $t^{\prime}-J$ model, invariant. The 
transformations of Eqs. \ref{i11} and \ref{i12}, which do leave $H_{\perp}$ 
invariant,
change the $d_{x^2-y^2}$
  symmetry of the pairs of Eq. \ref{tpgs} to $s$ and the $d_{x^2-y^2}$ 
singlet pairs of
  Eq. \ref{ccd} to extended-$s$ singlets and vice-versa.
  Hence, $d_{x^2-y^2}$ singlets of the form of Eq. \ref{ccd}
  are transformed to extended-$s$ singlets when the sign of $t^{\prime}$ is 
changed
  in the $t^{\prime}-J$ model, and vice-versa. More generally, operators are 
transformed
  according to Eq. \ref{i11} or \ref{i12} when the sign of $t^{\prime}$ is 
changed. (The
  two transformations give the same results for operators with the 
appropriate symmetries,
  which generally includes the operators of interest). This is different 
from ref. \onlinecite{martins1}.
  However, the main arguments and conclusions of that paper remain 
unchanged.

  \section{$\lowercase{t}^\prime -\lowercase{t}^{\prime 
\prime}-J_{\lowercase{z}} $ Model}

  For the more general
  $t^\prime -t^{\prime \prime}-J_z $ model,
  one obtains in lowest order the (unnormalized) wave functions
  \begin{eqnarray}
  |\psi^{\pm}_{k_{x},k_{y}}&>& \, = \,
  e^{-{ { \pi i k_{x} x } \over L }}
  (4 t^{\prime}) s_{x} s_{y} |h_{k_{x},k_{y}}> \nonumber\\
  && \, + \, e^{-{ { \pi i k_{y} y } \over L }}
  \Bigl[ (2 t^{\prime \prime})
  (s_{y}^{2} -  s_{x}^{2}) \pm \tau_{x,y} \Bigr]
    |v_{k_{x},k_{y}}>
  \label{ttwf1}
  \end{eqnarray}
  with energies
  \begin{eqnarray}
  \epsilon^{\pm}_{k_{x},k_{y}} =
  (-2 t^{\prime \prime}) (1 - s_{x}^{2} -  s_{y}^{2}) \pm \tau_{x,y} \, ,
  \label{tte}
  \end{eqnarray}
  where $s_{x} = \sin ( \pi k_{x}  / L )$,
  $s_{y} = \sin ( \pi k_{y} /  L )$, and
  \begin{eqnarray}
  \tau_{x,y} = 2 \Bigl\{ ( t^{\prime \prime} )^{2}
  ( s_{x}^{2} - s_{y}^{2} )^{2}
  + 4 ( t^{\prime} )^{2} s_{x}^{2} s_{y}^{2} \Bigr\} ^{ 1 \over 2}.
  \label{tau}
  \end{eqnarray}
  As a function of $t^{\prime}$ and $t^{\prime \prime}$, we find that
  the ground state symmetry of the pair is as shown in Fig. 1.

The $s$-wave and $d$-wave operators are of the form in Eq. \ref{tpgs}. The ground state associated 
with $p$-wave pairs is, in contrast, highly degenerate.
The multiple
  $p$-wave pair operators can be either $p_x$
  \begin{eqnarray}
  \Delta_{p_x}(k_y)={1 \over {L \sqrt 2}}
  \sum_{ x,y } e^{-{ { 2 \pi i k_{y} y } \over L }}
  a_{x,y} ( a_{x+1,y} - a_{x-1,y} )
  \label{px}
  \end{eqnarray}
  or $p_y$
  \begin{eqnarray}
  \Delta_{p_y}(k_x)={1 \over {L \sqrt 2}}
  \sum_{ x,y } e^{-{ { 2 \pi i k_{x} x } \over L }}
  a_{x,y} ( a_{x,y+1} - a_{x,y-1} ).
  \label{py}
  \end{eqnarray}
  In leading order, the $p_x$ states have energies independent of $k_y$, and the
  $p_y$ states have energies independent of $k_x$.
  Both $p$-wave pair operators change sign under a 180 degree rotation.

  \begin{figure}[h]
  \centering
  \mbox{}
  \includegraphics[width=6cm,angle=270]{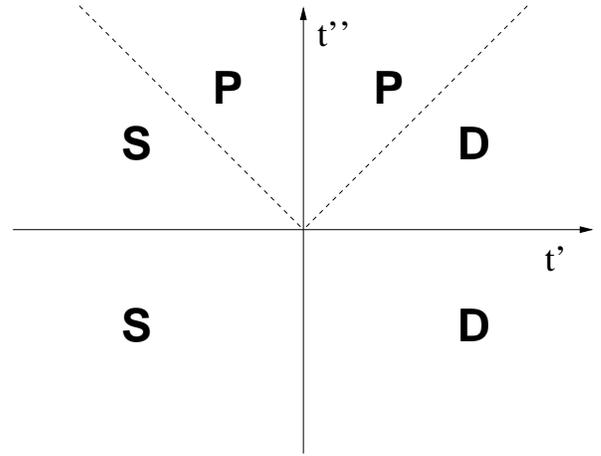}
  \vspace{0.1cm}
  \caption{\label{fig1}Hole pair symmetry in the strong-coupling limit of 
the
  $t^\prime - t^{\prime\prime}-J_z$ model
  as a function of $t^\prime$ and $t^{\prime\prime}$.
  The legend is as follows: ``$D$'' denotes $d_{x^{2} - y^{2}}$, ``$P$'' 
denotes
  $p_x$ or $p_y$, and ``$S$'' denotes extended-$s$. The axes cross
  at $t^{\prime}=0$ and $t^{\prime\prime}=0$.}
  \end{figure}

  \section{$\lowercase{t}-J_{\lowercase{z}}$ Model}

  We next consider the strong-coupling limit of the $t-J_z$ model.
  To lowest order, we find that
  this maps onto the above strong-coupling limit
  of the
  $t^{\prime} - t^{\prime \prime} - J_z$ model with
  \begin{eqnarray}
  t_{t,eff.}^{\prime} = t_{t,eff.}^{\prime \prime} =
  { 2 \over 3 }
  \left( { { t^{2} } \over { J_{z} } } \right).
  \label{teff}
  \end{eqnarray}
  From Eq. \ref{tte},
  the lower band of the pair excitation spectrum then
  becomes flat, with wave functions
  \begin{eqnarray}
  |\psi^{-}_{k_{x},k_{y}}> &=&
  {1 \over {\sqrt 2}} \,
  \Bigl\{ e^{-{ { \pi i k_{x} x } \over L }}
  \, s_{y} |h_{k_{x},k_{y}}> \nonumber\\
  && \, \, \, \, \, \, \, \,  - \,  \,
  e^{-{ { \pi i k_{y} y } \over L }}  \,
  s_{x} |v_{k_{x},k_{y}}> \, \Bigr\}.
  \label{twf}
  \end{eqnarray}
  Flat pair bands were
  also found \cite{shraiman,trugman}
  for related (though different) models and/or treatments.
  In ref. \onlinecite{cumulant},
  a five-fold degeneracy of strong-coupling
  $t-J_z$ pairs of pure
  $d$ or $p$ symmetry was noted.

  We see from Eq. \ref{teff}
  that, to lowest order, the strong-coupling
  $t-J_z$ model lies on the (rightmost)
  boundary in Fig. 1 between $d$-wave and $p$-wave symmetry,
  providing a simple picture for competition between these two states.
  In the next higher order, neglecting constant additive terms,
  the energies of the lower pair band separate into
  \begin{eqnarray}
  \epsilon^{-}_{k_{x},k_{y}} &=&
  \Bigl( - { 8 \over 45 } \Bigr)
  \Bigl( { { t^{4} } \over { J_{z}^3 } } \Bigr)
  \Bigl( 2 - c_x - c_y \Bigr)^{-1} \nonumber\\
  && \bigl\{ c_{x}^{2} + c_{y}^{2} + 4 c_{x} c_{y}
       - 31 c_{x} - 31 c_{y} + 56 \bigr\},
  \label{fourth}
  \end{eqnarray}
  where here $c_{x} = \cos ( 2 \pi k_{x} / L )$
  and
  $c_{y} = \cos ( 2 \pi k_{y} / L )$.
  Two-hole lower band dispersion curves were previously calculated
  numerically using a variational method \cite{eder92} and series
  expansions \cite{hamer98}.
  We then find (in agreement with refs. \onlinecite{hamer98} and \onlinecite{cumulant})
  that the pure $d$-wave ($t^{\prime} > 0$)
  state of Eq. \ref{tpgs}
  is selected as the lower pair band ground state.
  However, the closeness to $p$-wave symmetry may
  provide an explanation for
  the low-energy $p$-wave ``quasi-pair'' peaks seen numerically
  in small $t-J$ and $t-J_z$ clusters\cite{dagottop}.

  We also note that several different techniques have suggested that the
  symmetry of a doped hole pair in the $t-J_z$ and/or $t-J$ models may be
  $p$-wave for some range of generally intermediate or small $J_z/t$ or 
$J/t$
  \cite{hamer98,eder92,chernleung99,leung02}. However, this has not been
  rigorously confirmed one way or the other \cite{sorella,poilscal02}.

  One can
  perturbatively construct increasingly extended quasi-pair states
  for the $t-J_z$ model.
  Combining results for the NN $d$-wave pair operators
  for ground states $| \Phi_a>$ and $| \Phi_b>$,
  one finds the lowest order correction for the singlet pair
  operator of Eq. \ref{ccd}
  \begin{eqnarray}
  \Delta^{\prime}_d=\Bigl( - {2 \over 3} \Bigr)
  \Bigl( {t \over {J_z}} \Bigr)
  \Bigl( {1 \over L} \Bigr)
  \sum_{x,y,\sigma}
  \sigma  & \nonumber\\
  \Bigl\{ \bigl( c^{\dagger}_{x+1,y,\sigma} c_{x+1,y,-\sigma} -
  & c^{\dagger}_{x,y+1,-\sigma} c_{x,y+1,\sigma} \bigr) \nonumber\\
            c_{x+1,y+1,\sigma} c_{x,y,\sigma} \,\,\,\,\, & \nonumber\\
    \mbox{} + \bigl( c^{\dagger}_{x+1,y,\sigma} c_{x+1,y,-\sigma} -
  & c^{\dagger}_{x,y-1,-\sigma} c_{x,y-1,\sigma} \bigr) \nonumber\\
            c_{x+1,y-1,\sigma} c_{x,y,\sigma} \Big\}. &
  \label{dpert}
  \end{eqnarray}
  When operating on the appropriate N\'eel state,
  each of the above terms consists of a diagonal hole pair
  ``dressed'' with a singlet pair of electrons straddling the bond
  connecting the pair of holes, as was found in numerical
  $t-J$ simulations\cite{whitescal}. This form basically arises
  from the disruption in the local spin order when an electron of
  particular spin orientation and one hole of a NN hole pair exchange sites 
to form a diagonal hole pair.
  We note that the contribution
  from pairs a distance of
  two lattice sites apart, nominally also of
  order $t / J_z$,
  vanishes identically
  in this order. (Variational calculations found a reduction
  of such a contribution \cite{wrobel98}.)
  This vanishing may provide an explanation for why only
  NN and diagonal
  hole correlations appear to dominate in the $t-J$
  model near half filling for moderate to
  large $J / t$\cite{whitescal,didier}.

If one adds the necessary terms to the operator of Eq. \ref{dpert}
  to impose a type of spin rotational invariance, one obtains
  the composite pair operator invented in ref.  \onlinecite{didier}
  to give a diagonal singlet pair
  with $d_{x^2 - y^2}$ symmetry.
  The type of operator of ref. \onlinecite{didier} was also generated \cite{whitescal} in a
  $2\times2$ plaquette by applying $e^{-\tau H}$ to the pair operator
  of Eq. \ref{ccd}. 
  The operator of Eq. \ref{dpert}, which emerges naturally from
  perturbation theory, also has
  $d_{x^2 - y^2}$ symmetry.

  We also note that, since we calculate energy spectra and
  wave functions, our results and approach can be used to
  calculate finite-temperature and real frequency
  properties. Our results can also be easily extended to periodic
  $n$-leg ladders with even $n$.
  However, we do not pursue those issues here.

  We in addition note that, in contrast to Eqs. \ref{i11} through \ref{i14},
  the phase transformation that is equivalent to changing the sign of
  $t$ is given by
  \begin{eqnarray}
  d_{x,y,\sigma}=(-1)^{x+y}c_{x,y,\sigma}.
  \label{i15}
  \end{eqnarray}
  This transformation leaves the spin-spin interactions $H_0$ and
  the $H_{\perp}$ of Eq.  \ref{jperp} invariant. Under
  this transformation the symmetries of the hole pair operators of Eqs. 
\ref{tpgs},
  \ref{ccd}, and \ref{dpert} all remain the same.

\section {$\lowercase{t}-\lowercase{t}^{\prime}-\lowercase{t}^{\prime\prime}-J_{\lowercase{z}}$ 
Model and intermediate coupling}

  For $|t|$, $|t^{\prime}|$, $|t^{\prime\prime}|\ll J_z$;
  $|t| \, \gg \, |t^{\prime}|,|t^{\prime\prime}|$; and
  $|t^{\prime}/J_z| \, \gg \, (t/J_z)^4$ or
  $|t^{\prime\prime}/J_z| \, \gg \, (t/J_z)^4$
  (so that lowest order terms dominate),
  one can use Eq. \ref{teff} to define a
  \begin{eqnarray}
  t^{\prime}_{eff.}=t^{\prime}+t^{\prime}_{t,eff.}=t^{\prime}+{2 \over 3}
  {{t^2} \over {J_z}}
  \label{i20}
  \end{eqnarray}
  and
  \begin{eqnarray}
t^{\prime\prime}_{eff.}=t^{\prime\prime}+t^{\prime\prime}_{t,eff.}=t^{\prime\prime}+{2 
\over 3}
  {{t^2} \over {J_z}}.
  \label{i21}
  \end{eqnarray}
  One can then simply use Fig. 1 as a guide to pair symmetry.

  As one step towards investigating the further range of validity
  of our approach, we performed analytic calculations of the
  $t-t^{\prime}-J_{z}-J_{\perp}$ model
  (see Eq. \ref{jperp})
  on a $2\times2$ plaquette.
  In what follows, the four plaquette sites are consecutively numbered
  as one goes around the plaquette edge. Spin-spin interactions and
  $t$ hoppings are between consecutive sites, $t^{\prime}$ hoppings are
  between diagonal sites, and there are {\it no} periodic boundary 
conditions
  (though such boundary conditions would only renormalize parameter values).

  With no holes (four electrons, one per site), we found that the ground 
state had
  $d_{x^2-y^2}$ symmetry, as is the case for the $t-J$ model on a plaquette
  \cite{scaltrug,whitescal}, for all $J_z > 0$, $J_{\perp}\geq 0$. (There 
are
  two degenerate states only when $J_{\perp}=0$, both of $d$-wave symmetry).

  For two holes, when
  \begin{eqnarray}
  t^{\prime}>t^{\prime}_{cross.}={1 \over 8} \, 
[(J_z+J_{\perp})-\sqrt{(J_z+J_{\perp})^2
  +64t^2}]
  \label{pq1}
  \end{eqnarray}
  (the more ``physical'' parameter regime), the ground state energy is
  \begin{eqnarray}
  E_0=-{1 \over 2}[\alpha+(\alpha^2+32t^2)^{1 \over 2}],
  \label{pq2}
  \end{eqnarray}
  where
  \begin{eqnarray}
  \alpha={{J_z} \over 2}+{{J_{\perp}} \over 2}+2t^{\prime}.
  \label{pq3}
  \end{eqnarray}
  (We assume $J_z>0$, $J_{\perp}\geq0$, and either $t\neq 0$ or
  $t^{\prime}\neq 0$). Defining \cite{whitescal}
  \begin{eqnarray}
  \Delta^{\dagger}_{ij}={1 \over 
{\sqrt{2}}}(c^{\dagger}_{i\uparrow}c^{\dagger}_{j\downarrow}
  -c^{\dagger}_{i\downarrow}c^{\dagger}_{j\uparrow})
  \label{pq4}
  \end{eqnarray}
  with $\Delta^{\dagger}_{ij}=\Delta^{\dagger}_{ji}$, where $i$ and $j$ 
refer
  to plaquette sites, the ground state is a linear combination of the two 
states
  \begin{eqnarray}
  |\psi_1\rangle={1 \over 
{\sqrt{2}}}(\Delta^{\dagger}_{13}+\Delta^{\dagger}_{24})|0\rangle
  \label{pq5}
  \end{eqnarray}
  and
  \begin{eqnarray}
  |\psi_2\rangle={1 \over 2}(\Delta^{\dagger}_{12}+\Delta^{\dagger}_{23}+
  \Delta^{\dagger}_{34}+\Delta^{\dagger}_{41})|0\rangle.
  \label{pq6}
  \end{eqnarray}
  The ground state hence has $s$-wave symmetry, as was
  earlier found for the two-hole doped
  plaquette $t-J$ model \cite{scaltrug,whitescal}. As discussed previously
  \cite{scaltrug,whitescal}, this implies that the hole pair symmetry
  is $d$-wave. We note that with $t^{\prime}=0$, which automatically
  satisfies Eq. \ref{pq1}, the hole pair symmetry for the plaquette
  $t-J_z-J_{\perp}$ model is always $d-$wave, as in the strong-coupling
  limit of the general $t-J_z$ model. (However, because there are no lattice sites a distance of two lattice
spacings apart on the plaquette, for $t/J_{z} << 1$ the plaquette $t-J_z$  
model reduces to a $t^{\prime}-J_z$ model with $t^{\prime} > 0$ rather than a full
$t^{\prime}-t^{{\prime}{\prime}}-J_z$ model, enhancing $d$-wave pairing compared to a larger lattice.)

  However, there is a level crossing at $t^{\prime}=t^{\prime}_{cross.}$,
  and when $t^{\prime} < t^{\prime}_{cross.}$ the ground state has energy
  \begin{eqnarray}
  E_0=-{{J_z} \over {2}}-{{J_{\perp}} \over 2}+2t^{\prime}
  \label{pq7}
  \end{eqnarray}
  and wave function
  \begin{eqnarray}
  |\psi_3\rangle={1 \over {2}}(\Delta^{\dagger}_{12}-\Delta^{\dagger}_{23}+
  \Delta^{\dagger}_{34}-\Delta^{\dagger}_{41})|0\rangle.
  \label{pq8}
  \end{eqnarray}
  $|\psi_3\rangle$ has $d$-wave symmetry, implying that the hole pair
  symmetry is now $s$-wave. We note that with $t=0$, giving
  $t^{\prime}_{cross.}=0$, the $t^{\prime}-J_z-J_{\perp}$ model hole
  pair symmetry on a plaquette is $d$-wave for $t^{\prime}>0$ and $s$-wave
  for $t^{\prime} < 0$, as in the general $t^{\prime}-J_z$ model strong 
coupling
  case.

  As mentioned previously, some approaches have suggested $p$-wave
  hole-pair symmetry for some range of generally intermediate or
  small $J/t$ or $J_z/t$ \cite{hamer98,eder92,chernleung99,leung02}.
  However, other $t-J$ numerical results \cite{cumulant,sorella,poilscal02,didier}
  indicate that $d$-wave features of the strong-coupling limit may persist 
down to
  physically relevant intermediate coupling ($J/t\approx 0.3 - 0.5$).
  This and the above plaquette results are consistent with the strong-coupling
  $t- t^\prime -t^{\prime \prime}-J_z$ model as a potentially
  useful starting point to explore the more physically relevant 
intermediate-coupling
  $t- t^\prime -t^{\prime \prime}-J$ model.

  \begin{figure}[h]
  \centering
  \mbox{}
  \includegraphics[width=6cm,angle=270]{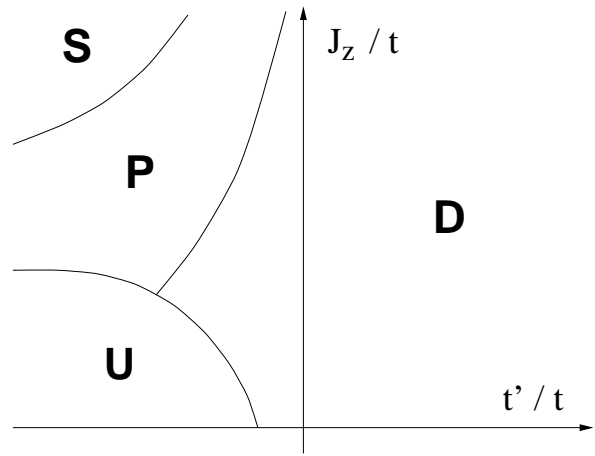}
  \vspace{0.1cm}
  \caption{\label{fig2}Qualitative diagram of proposed hole pair symmetry
  for the $t - t^\prime-J_z$ model.
  ``$D$'', ``$P$'',
  and ``$S$'' denote the same as in Fig. 1, and
  no prediction is made for region ``$U$''. Axes cross at $t^{\prime}/t=0$
  and intermediate $J_z/t$ ($\sim0.3 - 0.5$). The possibility of $p$-wave 
symmetry
  for $t^{\prime}=0$ and intermediate $J_z/t$ is not shown.}
  \end{figure}

  As a rough guide to the general effects of $t^{\prime}$ and 
$t^{\prime\prime}$
  on pairing symmetry, one can begin with a $d$-wave state in Fig. 1 arising
  from the $t$ term, presumably in the upper right quadrant. Assuming pairs of pure symmetry 
  and that strong coupling qualitatively extends to $|t^{\prime}|$, 
$|t^{\prime\prime}|\sim J,
  J_z$, both $t^{\prime}>0$ and $t^{\prime\prime}<0$ will tend to move
  one deeper into the $d$-wave region; however, $t^{\prime}<0$ and 
$t^{\prime\prime}>0$
  will tend to move one towards, and perhaps into, the $p$-wave region. 
Previous $t-J$
  work has argued and/or indicated numerically that $t^{\prime}>0$ 
strengthens
  $d$-wave pairing while $t^{\prime}<0$ weakens it 
\cite{whitescal1,martins1}, and
  it was found on a 32-site lattice that a particular $t^{\prime}<0$ and
  $t^{\prime\prime}>0$ together favored a $p$-wave pair \cite{leung02}. 
Also, if one
  starts with a possible $t-J_z$ or $t-J$ $p$-wave pair (upper right 
quadrant),
  $t^{\prime}<0$ and $t^{\prime\prime}>0$ would tend to move one deeper into 
the
  $p$-wave region while $t^{\prime}>0$ and $t^{\prime\prime}<0$ would tend 
to move
  one toward, and perhaps even into, the $d$-wave region.

  Based on this and our strong-coupling results
  (assuming pairs of pure symmetry),
  we show in Fig. 2 qualitative predictions of the hole pair
  symmetry for the $t - t^{\prime} - J_z$ model. We do not show
  in Fig. 2 the possibility of a cross-over to $p$-wave symmetry for
  the $t-J_z$ model mentioned above.
  We believe the predictions shown apply to the
  $t - t^{\prime} - J$ model as well,
  with a comparatively smaller $p$-wave region due to
  larger energy differences between $t-J$ $p$-wave and $d$-wave
  pair states \cite{dagottop}. An additional
  $t^{\prime \prime} > 0$ would tend to enlarge the $p$-wave region, 
and an additional $t^{\prime \prime} < 0$ would tend to enlarge the $d$-wave region.
 Note
  that in Fig. 2, the horizontal axis cuts the vertical axis
  at intermediate $J_z/t$ ($\sim0.3 - 0.5$).

  $t > 0$ for the hole-doped cuprates, and estimates
  for $t^{\prime}$ and $t^{\prime \prime}$
  are typically in the ranges
  $t^{\prime} \approx (-0.1)t - (-0.5)t$, and
  $t^{\prime \prime} \approx 0.0 - (0.3)t$, generally
  with  $|t^{\prime}|>|t^{\prime\prime}| \,$ \cite{param,onefit}.
  Both these signs of $t^{\prime}$ and $t^{\prime \prime}$ are those which
  could tend to drive the pair symmetry to $p$-wave,
  raising the issue of the hole pair
  symmetry in the intermediate-coupling regime. We note that $s$-wave is 
also possible,
  though we believe it less likely at intermediate coupling.

  It would be interesting to try to determine numerically
  whether the symmetry of two doped holes in the
  $t- t^\prime -t^{\prime \prime}-J$ model is in fact $d$-wave for
  the experimentally relevant values of
  $t, t^\prime, t^{\prime \prime}$, and $J$ (e.g., $J/t \approx 0.3 - 0.5$).
  However, drawing conclusions from exact diagonalization and other
  current approaches may be challenging (see, e.g., refs. 
\onlinecite{onefitsps}
  and \onlinecite{leung02}) due to the possibility of uncontrolled
  finite-size or other errors, and such approaches sometimes give 
conflicting results.
  If the symmetry
  were established to be $p$-wave rather than $d$-wave,
  it would suggest that the
  $t- t^\prime -t^{\prime \prime}-J$ model
  by itself could be incomplete
  as a model for high-$T_c$ superconductivity.
  In that case,
  one possibility for restoring $d$-wave symmetry
  could be the addition of electron-phonon coupling in the
  $d$-channel \cite{phonon}.
  In either case,
  it may also be of interest to explore how the existence
  of or nearness to $p$-wave symmetry, which effectively reduces the
  dimensionality of the hole pair wave function from 2D to 1D,
  would correlate with possible stripe phases observed
  experimentally \cite{tranquada,striping} and in numerical
  simulations \cite{whitescal00,kampfscal01,whitescal03}.

  Another interesting possibility is the existence of a
  novel exotic state which may be close in energy to the
  $d_{x^2-y^2}$ superconducting cuprate ground state.
  Such a {\it competing phases} scenario \cite{davis-elbio} could become 
even more involved
  if one takes in account recent numerical calculations \cite{martins-ferro}
  pointing to the existence of strong ferromagnetic fluctuations in the 
vicinity of
  doped holes for realistic values of $t^\prime < 0$.

  \section{Summary}

  In summary, we have investigated analytically
  the ground state pair symmetry and excitation
  spectra of two holes doped into the half-filled
  $t- t^\prime -t^{\prime \prime}-J_z $ model
  in the strong-coupling
  limit.
  In lowest order, this reduces to considering the
  $t^{\prime}- t^{\prime \prime} -J_z $ model, where we found
  regions of ground state $d$-wave, $s$-wave, and (degenerate) $p$-wave symmetry,
  depending upon the signs and relative magnitudes of $t^\prime$
  and $t^{\prime \prime}$.
  We next found that the $t-J_z$ model in lowest order was
  on the boundary between $d$-wave and $p$-wave pair symmetry,
  with a flat lower pair dispersion, providing a simple picture
  for the competition between $d$ and $p$ symmetries.
  In higher order, $d$-wave symmetry was selected from the lower pair band.
  However, because of the closeness to $p$-wave symmetry, we predict that 
the appropriate
  $t^\prime < 0$ and/or $t^{\prime \prime} > 0$ added to the $t-J_z$ or 
$t-J$
  models with intermediate to large $J_z$ or $J$
  should drive them into $p$-wave pairs,  and perhaps even $p$-wave 
superconductivity.
   These signs of $ t^\prime$ and $ t^{\prime \prime}$ are those found
   in the hole-doped cuprates. (In contrast, $t^{\prime}>0$ and/or 
$ t^{\prime \prime} < 0$ tend to promote $d$-wave symmetry.)
  This $p$-wave tendency could be strengthened following results which 
suggest
  $p$-wave pair symmetry for intermediate or small $J/t$ or $J_z/t$ in the
  $t-J$ or $t-J_z$ models \cite{hamer98,eder92,chernleung99,leung02},
  though such results have not been rigorously confirmed 
\cite{sorella,poilscal02}.

We constructed a perturbative correction to the
  nearest-neighbor $d$-wave pair, giving a more extended quasi-pair, and found that it was
  similar to the $d$-wave
  composite operator invented in ref. \onlinecite{didier} and qualitatively 
consistent
  with previous numerical results \cite{whitescal}.  The quasi-pair included 
a
  contribution from sites $\sqrt{2}$ lattice spacings apart, but the 
same-order contribution
  from sites $2$ lattice spacings apart vanished identically.  The structure 
of the quasi-pair
  derived from the disruption in local spin order under the exchange of one of 
the nearest neighbor holes and an electron.

  We explored ranges of validity of the perturbative approach of this paper 
using
  a $2\times2$ plaquette and results from other work 
\cite{cumulant,sorella,poilscal02,didier}.
  Lastly, we discussed implications for
  the experimentally relevant parameter regime. These included
  the possibility of $p$-wave symmetry for two doped holes, which would
  suggest  that the
  $t - t^{\prime}- t^{\prime \prime} -J$ model could be incomplete as
  a high-$T_c$ model, or perhaps a phase competition scenario
  between $d$-wave superconductivity and a $p$-wave state.

  R. M. Fye would like to thank R. Duncan for his hospitality at UNM in 
1996-1997,
  when his part of the research was performed.
  G. Martins and E. Dagotto are supported by NSF grants
  DMR-0122523 and DMR-0303348. Additional support from
  Martech (FSU) is also acknowledged.


\begin{thebibliography}{natbib}

  \bibitem{dexp1} D. J. Van Harlingen,
  Rev. Mod Phys. {\bf 67}, 515 (1995).

  \bibitem{annet} J. F. Annett, N. Goldenfield, and A. J. Leggett,
  in {\it Physical Properties of High Temperature
  Superconductors}, Vol. V, ed. D.M. Ginsberg (1996).

  \bibitem{tsuei-kirtley} C. C. Tsuei and J. R. Kirtley,
  Rev. Mod. Phys. {\bf 72}, 969 (2000).

  \bibitem{dagottorev} E. Dagotto,
  Rev. Mod Phys. {\bf 66}, 763 (1994).

  \bibitem{schrieffer} J. R. Schrieffer,
  Solid State Commun. {\bf 92}, 129 (1994).

  \bibitem{reviews} A. P. Kampf, Phys. Rep. {\bf 249},
  219 (1994); W. Brenig, Phys. Rep. {\bf 251}, 155 (1995);
  and refs. therein.

  \bibitem{scalrev} D. J. Scalapino,
  Phys. Rep. {\bf 250}, 329 (1995).

\bibitem{joynt} B. E. C. Koltenbah and R. Joynt, Rep. Prog. Phys. {\bf 60}, 23 (1997).

\bibitem{hlubina} R. Hlubina, Phys. Rev. B {\bf 59}, 9600 (1999).

  \bibitem{scaltrug} D. J. Scalapino and S.A. Trugman,
  Phil. Mag. B {\bf 74}, 607 (1996).

  \bibitem{whitescal} S.R. White and D. J. Scalapino,
  Phys. Rev. B {\bf 55}, 6504 (1997). See also A. Moreo and
  E. Dagotto, Phys. Rev. B {\bf 41}, 9488 (1990).

  \bibitem{pseudo1} R.S. Markiewicz, J. Phys. Chem. Sol. {\bf 58}, 1179 
(1997),
  and refs. therein.

  \bibitem{pseudo2} A.V. Puchkov, D. N. Basov, and T. Timusk, J. Phys.-Cond. 
Matt. {\bf 8}, 10049 (1996).

  \bibitem{preform} J. M. Harris, Z. -X. Shen, P. J. White, D. S. Marshall, 
M. C. Schabel,
  J. N. Eckstein and I. Bozovic, Phys. Rev. B {\bf 54}, 15665 (1996); J. M. 
Harris, P. J. White,
  Z.-X. Shen, H. Ikeda, R. Yoshizaki, H. Eisaki, S. Uchida, W. D. Si, J. W. 
Xiong, Z.-X. Zhao, and D. S. Dessau,
  Phys. Rev. Lett. {\bf 79}, 143 (1997).

  \bibitem{timusk-statt} T. Timusk and B. Statt,
  Rep. Prog. Phys. {\bf 62}, 61 (1999).

  \bibitem{tallon} J. L. Tallon and J. W. Loram, Physica C {\bf 349}, 53 
(2001).

  \bibitem{nan} A. Paramekanti, M. Randeria and N. Trivedi,
  Phys. Rev. Lett. {\bf 87}, 217002 (2001); {\it idem},
  cond-mat/0305611.

  \bibitem{dagottop} E. Dagotto, J. Riera and A. P. Young,
  Phys. Rev. B {\bf 42}, 2347 (1990).

  \bibitem{tjz}
  Z. Liu and E. Manousakis,
  Phys. Rev. B {\bf 45}, 2425 (1992).

  \bibitem{hamer98} C. J. Hamer, Z. Weihong, and J. Oitmaa, Phys. Rev. B 
{\bf 58}, 15508 (1998).

\bibitem{boninsegni} Some different behaviors in certain of the $t-J$ and $t-J_z$ 
model properties were suggested in M. Boninsegni, Phys. Rev. Lett. {\bf 87}, 087201 (2001). 
However, these behaviors are suggested to be similar in, e. g., 
A. L. Chernyshev, S. R. White and A. H. Castro Neto, Phys. Rev. B {\bf 65}, 
214527 (2002).

  \bibitem{cumulant} P. Prelov\u {s}ek, I. Sega, and J. Bon\u{c}a,
  Phys. Rev. B {\bf 42}, 10706 (1990).

  \bibitem{param}
  H. Eskes, G. A. Sawatzky, and L. F. Feiner, Physica C {\bf 160}, 424 
(1989);
  M. S. Hybertsen, E. B. Stechel, M. Schluter, and D. R. Jennison,
  Phys. Rev. B {\bf 41}, 11068 (1990);
  T. Tohyama and S. Maekawa, J. Phys. Soc. Jpn. {\bf 59}, 1760 (1990);
  D.C.  Mattis and J. M.  Wheatley,
  Mod.  Phys.  Lett.  {\bf 9}, {1107} (1995);
  V.I. Belinicher, A. L. Chernyshev, and V. A. Shubin,
  Phys. Rev. B {\bf 53}, 335 (1996);
  L.F. Feiner, J. H. Jefferson and R. Raimondi,
  Phys. Rev. B {\bf 53}, 8751 (1996);
  R. Hayn, A. F. Barabanov, J. Schulenburg, Zeit. Phys. B {\bf 102}, 359 
(1997);
  E. Pavarini, I. Dasgupta, T. Saha-Dasgupta, O. Jepsen, and O.K. Andersen,
  Phys. Rev. Lett. {\bf 87}, 047003 (2001).

  \bibitem{onefit}
  A. Nazarenko, K. J. E. Vos, S. Haas, E. Dagotto, and R. J. Gooding,
  Phys. Rev. B {\bf 51}, 8676 (1995);
  P. W. Leung and R. J. Gooding,
  Phys. Rev. B {\bf 52}, 15711 (1995);
  B. O. Wells, Z. -X. Shen, A. Matsuura, D. M. King, M. A. Kastner, M. 
Greven, and R. J. Birgeneau,
  Phys. Rev. Lett. {\bf 74}, 964 (1995);
  L.F. Feiner, J. H. Jefferson, and R. Raimondi,
  Phys. Rev. Lett. {\bf 76}, 4939 (1996);
  T. Xiang and J. M. Wheatley,
  Phys. Rev. B {\bf 54}, 12653 (1996);
  V.I. Belinicher, A. L. Chernyshev, and V. A. Shubin,
  Phys. Rev. B {\bf 54}, 14914 (1996);
  D. Duffy, A. Nazarenko, S. Haas, A. Moreo, J. Riera, and E. Dagotto, Phys. 
Rev. B {\bf 56}, 5597 (1997);
  C. Kim, P. J. White, Z.-X. Shen, T. Tohyama, Y. Shibata, S. Maekawa, B. O. 
Wells,, Y. J. Kim, R.
  J. Birgeneau, and M. A. Kastner, Phys. Rev. Lett. {\bf 80}, 4245 (1998);
  T. Tohyama and S. Maekawa, Superc. Sc. Techn. {\bf 13},
  R17 (2000), and refs. therein; F. Ronning,  C. Kim, K. M. Shen, N. P. 
Armitage, A. Damascelli,
  D. H. Lu, D. L. Feng, Z.-X. Shen, L. L. Miller, Y.-J. Kim, F. Chou, and I. 
Terasaki, Phys. Rev. B
  {\bf 67}, 035113 (2003);
  A. Damascelli, Z. Hussain, and Z.-X. Shen, Rev. Mod. Phys. {\bf 75}, 473 
(2003).

  \bibitem{onefitsps} R. Eder, Y. Ohta, and
  G.A. Sawatzky,
  Phys. Rev. B {\bf 55}, 3414 (1997).

  \bibitem{whitescal1} S. R. White and D. J. Scalapino, Phys. Rev. B
  {\bf 60}, R753 (1999).

  \bibitem{martins1} G. B. Martins, J. C. Xavier, L. Arrachea, and E. 
Dagotto, Phys. Rev. B
  {\bf 64}, 180513 (2001).

  \bibitem{afvh} E. Dagotto, A. Nazarenko and A. Moreo,
  Phys. Rev. Lett. {\bf 74}, 310 (1995).

  \bibitem{shraiman} B.I. Shraiman and E.D. Siggia,
  Phys. Rev. Lett. {\bf 60}, 740 (1988).

  \bibitem{trugman} S.A. Trugman,
  Phys. Rev. B {\bf 37}, 1597 (1988).

  \bibitem{eder92} R. Eder, Phys. Rev. B {\bf 45}, 319 (1992).

  \bibitem{chernleung99} A. L. Chernyshev and P. W. Leung, Phys. Rev. B {\bf 
60},
  1592 (1999).

  \bibitem{leung02} P. W. Leung, Phys. Rev. B {\bf 65}, 205101 (2002).

  \bibitem{sorella} S. Sorella, G. B. Martins, F. Becca, C. Gazza, L. 
Capriotti,
  A. Parola, and E. Dagotto, Phys. Rev. Lett. {\bf 88} 117002 (2002).

  \bibitem{poilscal02} D. Poilblanc and D. J. Scalapino, Phys. Rev. B {\bf 
66}, 052513 (2002).

  \bibitem{wrobel98} P. W. Wrobel and R. Eder, Phys. Rev. B {\bf 58}, 15160 
(1998).

  \bibitem{didier} D. Poilblanc,
  Phys. Rev. B {\bf 49}, 1477 (1994).

  \bibitem{phonon} O. K. Andersen, S. Y. Savrasov, O. Jepsen, A. I. 
Liechtenstein,
  J. Low temp. Phys. {\bf 105}, 285 (1996). See also A. Nazarenko and 
E. Dagotto, Phys. Rev. B {\bf 53}, R2987 (1996).

  \bibitem{tranquada}
  J. M. Tranquada, J. Phys. Chem. Solids {\bf 60}, 1019 (1999).

  \bibitem{striping} A. Bianconi, N. L. Saini, A. Lanzara, M. Missori, T. 
Rossetti,
  H. Oyanagi, H. Yamaguchi, K. Oka, and T. Ito,
  Phys. Rev. Lett. {\bf 76}, 3412 (1996);
  J. M. Tranquada, J. D. Axe, N. Ichikawa, A. R. Moodenbaugh, Y. Nakamura, and S. 
Uchida,
  Phys. Rev. Lett. {\bf 78}, 338 (1997); and
  refs. therein.

  \bibitem{whitescal00} S. R. White and D. J. Scalapino, cond-mat/0006071 
(unpublished).

  \bibitem{kampfscal01} A. P. Kampf, D. J. Scalapino, and S. R. White, Phys. 
Rev. B {\bf 64}, 052509 (2001).

  \bibitem{whitescal03} S. R. White and D. J. Scalapino, cond-mat/0306545.

  \bibitem{davis-elbio} J. E. Hoffman, E. W. Hudson, K. M. Lang, V. 
Madhavan, H. Eisaki, S. Uchida, J. C. Davis, Science {\bf 295},
  466 (2002). See also E. Dagotto, {\it Nanoscale Phase Separation
  and Colossal Magnetoresistance}, Springer-Verlag, Berlin, 2002.

  \bibitem{martins-ferro} G. B. Martins and E. Dagotto,
  unpublished.

  \end{thebibliography}
  \end{document}